\documentclass{article}

\usepackage{arxiv}
\usepackage{multirow, makecell}
\usepackage{amsmath}
\usepackage{algorithm} 
\usepackage{algpseudocode}

\usepackage[utf8]{inputenc} 
\usepackage[T1]{fontenc}    
\usepackage{hyperref}       
\usepackage{url}            
\usepackage{booktabs}       
\usepackage{amsfonts}       
\usepackage{nicefrac}       
\usepackage{microtype}      
\usepackage{lipsum}
\usepackage{graphicx}
\usepackage{subcaption}
\graphicspath{ {./images/} }

\newcommand{\xobs}{X_\text{obs}}

\title{Generalised Bayes Updates with $f$-divergences through Probabilistic Classifiers}

\author{
 Owen Thomas \\
  Department of Biostatistics\\
  University of Oslo\\
  Oslo, Norway \\
  \texttt{o.m.t.thomas@medisin.uio.no} \\
  \And
 Henri Pesonen \\
  Department of Biostatistics\\
  University of Oslo\\
  Oslo, Norway \\
  \texttt{henri.pesonen@medisin.uio.no} \\
  \And
 Jukka Corander \\
  Department of Biostatistics\\
  University of Oslo\\
  Oslo, Norway \\
  \texttt{jukka.corander@medisin.uio.no} \\
}

\begin{document}
\maketitle
\begin{abstract}
A stream of algorithmic advances has steadily increased the popularity of the Bayesian approach as an inference paradigm, both from the theoretical and applied perspective. Even with apparent successes in numerous application fields, a rising concern is the robustness of Bayesian inference in the presence of model misspecification, which may lead to undesirable extreme behavior of the posterior distributions for large sample sizes. Generalized belief updating with a loss function represents a central principle to making Bayesian inference more robust and less vulnerable to deviations from the assumed model. Here we consider such updates with $f$-divergences to quantify a discrepancy between the assumed statistical model and the probability distribution which generated the observed data. Since the latter is generally unknown, estimation of the divergence may be viewed as an intractable problem. We show that the divergence becomes accessible through the use of probabilistic classifiers that can leverage an estimate of the ratio of two probability distributions even when one or both of them is unknown. We demonstrate the behavior of generalized belief updates for various specific choices under the $f$-divergence family. We show that for specific divergence functions such an approach can even improve on methods evaluating the correct model likelihood function analytically. 
\end{abstract}


\section{Introduction}

Bayesian reasoning offers a principled framework for probabilistic statistical inference and decision-making in the presence of uncertainty \cite{bernardo2009bayesian}. It is traditionally stated in the form of Bayes' theorem, in terms of the prior belief $p(\theta)$, posterior belief $p(\theta|\xobs)$, and the likelihood $p(\xobs|\theta)$ concerning a parameter $\theta$ and observed data $\xobs$:

\begin{align}\label{eqn:bayestheorem}
    p(\theta|\xobs)\propto p(\xobs|\theta)p(\theta)
\end{align}

However, Bayesian updates can exhibit undesirable behaviour in the presence of substantial model misspecification, i.e.~when the process that generated the data does not fall within the class of models contained in the prior.

Such behaviour can take the form of the posterior distribution concentrating onto the parameter value corresponding to the statistical model closest in Kullback-Leibler (KL) divergence to the unobserved true data-generating process $q_{\mathcal{T}}()$, which may result in a very confident belief in what may be a poor model if all the models in prior are poor \cite{walker2013bayesian}. Such behaviour has been observed and analysed in practical applications, especially thoroughly in the problem of phylogeny tree estimation \cite{yang2007fair,yang2018bayesian}. The KL divergence between the model likelihood $p(|\theta)$ and unobserved generative process $q_{\mathcal{T}}()$ is defined as follows:

\begin{align}
      D_{KL}\big(p(|\theta)||q_{\mathcal{T}}()\big) =  \int \log \frac{p(X|\theta)}{q_{\mathcal{T}}(X)} q_{\mathcal{T}}(X) dX
\end{align}

The presence of $q_{\mathcal{T}}(X)$ in the denominator of the ratio in the definition of the KL divergence can lead to the tails of the distributions being given large influence over the value of the divergence, when the statistician is often more interested in successfully capturing the features associated with the bulk of the probability mass. In the non-asymptotic domain, the inference can also be rendered unstable by misspecification, for example when the tails of the statistical model are overly light, giving undue influence to the outlying observed data points.

Various methods have been introduced to tackle misspecification, while retaining the desirable features of Bayesian inference. Generalised belief updates using generic loss functions $l(X,\theta)$ and tempering factor $w$ introduce the possibility of defining an alternative method for defining the influence of the data on the posterior belief while retaining the desirable Bayesian principle of coherency \cite{bissiri2016general}:

\begin{align}\label{eqn:loss_update}
    p(\theta|\xobs)\propto \exp\big(-w l(\xobs,\theta)\big) p(\theta)
\end{align}

If the negative log-likelihood (or the KL divergence, to within an additive constant) is used as a loss and the tempering $w$ is equal to one, familiar Bayes theorem from Equation \eqref{eqn:bayestheorem} is recovered. However, alternative choices of loss function and tempering are possible, including those defined analogously to discrepancies in the context of likelihood-free inference \cite{thomas2020split}, or statistical divergences other than the KL \cite{jewson2018principles}.

The use of different $f$-divergences is explored in \cite{jewson2018principles}, which presents derivations of the generalised belief distributions when using non-KL divergences as losses: the generalised posteriors are presented as functions of the statistical likelihood $p(X|\theta)$ and  the likelihood $g(X)$ of a model representing the statistician's best representation of their true beliefs concerning the distribution of the data. The choice of $f$-divergence is considered a modelling decision to be made according the subjective judgement of the statistician: each divergence gives influence to different features of the data generating distribution, resulting in distinct generalised belief updates. Total Variation Distance (TVD) divergences are recommended when the statistician wants their analysis to be robust to outliers in the data; the KL divergence is appropriate when the statistician wants to give the tails of the distributions a large influence on the belief update and make efficient use of a small amount of data; the alpha divergence represents a trade-off between the former and the latter when the parameter $\alpha$ increases from 0.5 to 1. Such an approach is tractable, but depends on additional generative modelling of the distribution used to represent the best accessible representation $g()$ of the data generating process. This is done nonparametrically using Kernel Density Estimation (KDE) in \cite{jewson2018principles}, but the additional task of performing generative modelling is acknowledged as a challenge that would appreciate further work.

An alternative to generative modelling for density estimation has been pursued via ratio estimation through probabilistic classification \cite{sugiyama2012density}. This is performed by training a probabilistic classifier to discriminate between samples from the numerator and denominator of a ratio, after which the probabilistic prediction of the classifier on the data of interest can be used to give a principled estimation of the ratio itself, without any generative modelling of either of the components of the ratio. 

The challenge of characterising densities from samples has been central to likelihood-free inference, when the likelihood function $p(X|\theta)$ cannot itself be evaluated \cite{sisson2018handbook}. In this context sample-based ratio estimation is tractable as it avoids the need to specify the likelihood function and can be performed just using samples drawn from a simulator. Ratio estimation has been used in a likelihood-free context for defining discrepancies, \cite{gutmann2018likelihood}, performing Bayesian updates \cite{thomas2016likelihood,hermans2019likelihood}, hypothesis testing \cite{cranmer2015approximating}, variational inference \cite{tran2017hierarchical,louppe2019adversarial}, and misspecficiation analysis \cite{thomas2019diagnosing}. Such methodology is also central to the methodology of Generative Adversarial Networks \cite{goodfellow2014generative}, in which a generative neural network is trained using the error rate of a discriminative neural network to train the generative model. The use of different $f$-divergences has been pursued with a neural network generative model \cite{nowozin2016f}, but to our knowledge has not been pursued in the context of model-based interpretable statistical inference.

The layout of this article as as follows: Section \ref{sec:methods} details the methodology necessary to perform efficient updates with $f$-divergence losses using classifiers and simulations. Section \ref{sec:examples} details the behaviour of such methods compared to methods using generative likelihood-based modelling approach presented in \cite{jewson2018principles}. Section \ref{sec:conclusion} concludes the article and discusses the results and possible future work.

\section{Methods}\label{sec:methods}

In this work we explore the use of classifiers and ratio-estimation to approximate the generalised posteriors based on $f$-divergences presented in \cite{jewson2018principles}. All of the generalised posteriors presented in \cite{jewson2018principles} are functions of the ratio $p(X|\theta)/g(X)$ between the statistical model likelihood and the generative process likelihood, but their work uses separate generative models with tractable likelihoods to characterise $p(X|\theta)$ and $g(X)$ separately.

The $f$-divergences $D_f\big(p(|\theta)||g()\big)$ can be defined as a general function $f$ of the ratio of two distributions $p(|\theta)/g()$ integrated with respect to the distribution $g()$:

\begin{align}\label{eqn:fdivergence}
    D_f\big(p(|\theta)||g()\big) = \int f\left(\frac{p(X|\theta)}{g(X)}\right)g(X) dX
\end{align}

The fact that $f$-divergences are solely functions of the ratio $p(X|\theta)/g(X)$ lends them naturally to approximation by discriminative ratio estimation. The work of performing generative modelling to establish $g(X)$ is thus avoided by training a classifier to discriminate between samples from the statistical model and observed data, and then using the classifier to approximate the log ratio $\log \big(p(X|\theta)/g(X)\big)$ directly. Pseudocode demonstrating how classifiers can be used to generate an approximation $\rho(\theta)$ to the log ratio $\log \big(p(X|\theta)/g(X)\big)$ is shown in Algorithm \ref{alg:forward_pseudo}.

The continuous integral in Equation \ref{eqn:fdivergence} can be approximated through evaluation of a sum of s function the ratio on a set of samples drawn from $g()$. If we interpret $g()$ to signify the statistician's best approximation of the true generative process, then it is appropriate to use the observed data $X_{obs}$ to represent samples drawn from $g(X)$, i.e.:

\begin{align}\label{eqn:forward_obs}
    D_f\big(p(|\theta)||g()\big) \approx \frac{1}{n_\text{obs}} \sum_{i=1}^{n_\text{obs}} f\left(\frac{p(\xobs^{i}|\theta)}{g(\xobs^{i})}\right), \quad \xobs^i\sim g(X)
\end{align}

The discrete approximation of the continuous integral from Equation \ref{eqn:forward_obs} is then used to define a belief update using each divergence as a loss function:

\begin{align}
    p(\theta|X_\text{obs},g,f) \propto \exp\Big(-n_\text{obs} D_f\big(p(|\theta)||g()\big)\Big) p(\theta)
\end{align}

We are implicitly using a tempering factor $w=1$: finding values of $w$ appropriate for a given belief update is an important statistical question, but is not the focus of this work.

\begin{algorithm}
	\caption{Pseudocode returning $\rho$ as an approximation to $\log\big( p(\xobs|\theta)/g(\xobs)\big)$ using a classifier $\mathcal{C}$}
	\begin{algorithmic}[1]
	    \State $\rho=0$
	    \State $n_\text{sim}=n_\text{obs}(1-K)/K$
	    \State Simulate $n_\text{sim}$ $X_\text{sim}\sim p(X|\theta)$
		\For {$k=1,2,\ldots K$ }
		    \State Partition observed data into fraction $1/K$ and $1-1/K$: $\xobs^{(k)}$, $\xobs^{(-k)}$
		    \State Calculate summaries $\phi_\text{sim}=\phi(X_\text{sim})$, $\phi_\text{obs}^{(k)}=\phi(\xobs^{(k)})$ and $\phi_\text{obs}^{(-k)}=\phi(\xobs^{(-k)})$
		    \State Create a vector of binary labels $y$ of $n_\text{sim}$ zeros and $n_\text{sim}$ ones.
		    \State Train a logistic classifier $\mathcal{C}$ to discriminate between $\phi_\text{sim}$ and $\phi_\text{obs}^{(-k)}$ using likelihood $p(y|\phi_\text{sim},\phi_\text{obs}^{(-k)},\mathcal{C})$
		    \State Predict the classifier decision function $\delta$ at $\phi_\text{obs}^{(k)}$: $\delta^{(k)}=\delta(\phi_\text{obs}^{(k)}|\phi_\text{sim},\phi_\text{obs}^{(-k)},\mathcal{C})$
		    \State $\rho=\rho+\sum(\delta^{(k)})/n_{obs}$
		\EndFor
		\State return $\rho$
	\end{algorithmic}\label{alg:forward_pseudo}
\end{algorithm}

We use the pointwise-estimates $\rho_i$ of the log ratio $ \log\big( p(X|\theta)/g(X)\big)$ evaluated at corresponding parameter values $\theta_i$ provided by Algorithm \ref{alg:forward_pseudo}, considering the value of the log ratio as a function $\boldsymbol\rho(\theta)$ of $\theta$. A Gaussian Process (GP) model \cite{williams2006gaussian} is used to model the log ratio as a nonparametric function of $\theta$:

\begin{align}
    \boldsymbol\rho(\theta) \sim \mathcal{GP}\big(0,K(\theta,\theta')\big)
\end{align}

A GP can both smooth between evaluations and also perform Bayesian Optimisation (BayesOpt) \cite{shahriari2015taking} to make efficient acquisitions informative to the mode of the log ratio, in the style of previous work on likelihood-free inference \cite{gutmann2016bayesian}.

The GP model uses mean values $\rho_i$ of the approximated log ratios as response variables and corresponding parameter values $\theta_i$ as covariates, transform through a kernel function $k(\theta,\theta')$. The predictive distribution as parameter value $\theta^*$ provides a mean $\mu(\theta^*)=\mathbb{E}(\boldsymbol\rho(\theta^*)|\theta^{*},\theta,\rho)$ and variance $\sigma^2=(\theta^*)\mathbb{V}(\boldsymbol\rho(\theta^*)|\theta^{*},\theta,\rho)$: these are used to construct an upper confidence bound acquisition function to select new values of $\theta$ that would be informative towards the mode of the function $\boldsymbol\rho(\theta)$. Upper confidence bound acquisition functions generally take the form $\mu(\theta^*)+\beta \sigma(\theta^*)$, where $\beta$ is a parameter defining the degree of exploration desired: the maximum of the acquisition function represents the tradeoff between exploring uncertainty in $\boldsymbol\rho(\theta)$ and exploiting large values of $\boldsymbol\rho(\theta)$, optimal for a given value of $\beta$. The acquisition function and kernel hyperparameters are updated dynamically as further evaluations are acquired as the optimisation proceeds: BayesOpt acquisitions are designed to providing information relevant to the global structure of the function, but also dense acquisitions near the mode, providing detailed information regarding the largest values.

\begin{figure}[h]

\begin{subfigure}{0.5\textwidth}
\includegraphics[width=0.9\linewidth, height=5cm]{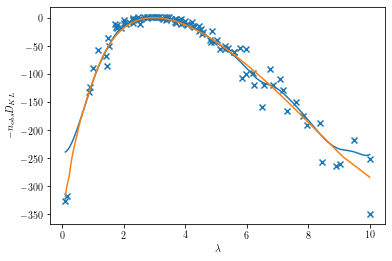} 
\caption{Variational Bayes Classifier}
\label{fig:subim1}
\end{subfigure}
\begin{subfigure}{0.5\textwidth}
\includegraphics[width=0.9\linewidth, height=5cm]{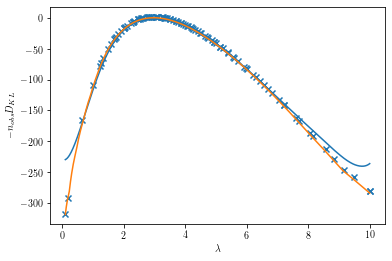}
\caption{KDE generative modelling}
\label{fig:subim2}
\end{subfigure}
\caption{BayesOpt acquisitions and GP predictive means in blue of the KL divergences using a variational Bayes classifier and likelihood-based KDE generative method, with the ratio using the unobserved true process in orange. The statistical model $p(X|\lambda)$ is a Poisson distribution $\mathrm{Pois}(\lambda)$, with 90 observed data points drawn from $q_{\mathcal{T}}(X)$, a Poisson distribution $\mathrm{Pois}(3)$}
\label{fig:acquisitions}
\end{figure}

The log-ratio is in principle a well-behaved function of the parameter space, but the variation associated with the finite number of simulations and pointwise classifier inference means that the individual evaluations can be considered noisy samples from the true log ratio values. As such, a Gaussian Process regression model with noise is appropriate to model the underlying smooth function.

We consider eight different divergences from which to construct generalised belief distributions: Kullback-Leibler, Squared Hellinger, TVD, and alpha divergences with $\alpha=[0.5,0.6,0.7,0.8,0.9]$, as defined in Table \ref{tab:divergences}.  An update using the KL as a loss is equivalent to a standard Bayesian update, with the influence of the generative distribution $g()$ being entirely absorbed into the normalisation: for this distribution we would assume the generative modelling method that assumes the statistical likelihood $f(|\theta)$ to match the update conditional on the true data generating process, assuming numerical stability, minimal influence of ratio truncation, and the success of the BayesOpt procedure.

Alpha divergences with $\alpha=0.5$ are equivalent the the Squared Hellinger distance with additional tempering of 0.25, which gives equal influence to the distributions $f(|\theta)$ and $g()$, while $\alpha=1.0$ is identical to the KL divergence, which does not use $g()$ in the belief update. Values of $\alpha$ increasing from 0.5 to 1.0 represent the increasing influence of $f(|\theta)$ relative to $g()$: we would expect a model assuming the statistical likelihood $f(|\theta)$ to perform increasingly effectively for increasing values of $\alpha$.

Some truncation of the ratio estimates was necessary to ensure stability, as the classifiers could sometimes return extreme values that substantially skew the mean value of the log ratio. Consequently, the predictive means of the Gaussian Process trained on the log ratios were truncated before transformation to $f$-divergences: the mean predictives on individual data points with values greater than 3 were set to 3, and those lower than -5 were set to -5. When estimating the TVD, the log ratios above zero were set to zero and those below -5 were set to -5.

\begin{table}
    \centering
    \begin{tabular}{c|c|c}
     Name & Divergence & Definition \\
     \hline
     Kullback-Leibler&  $D_{KL}\big(p(|\theta)||g()\big)$  & $\int \log \frac{p(X|\theta)}{g(X)} p(X|\theta) dX$ \\
      Squared Hellinger& $D_{Hell}^2\big(p(|\theta)||g()\big)$  & $1-\int \sqrt{p(X|\theta)g(X)}dX$ \\
    Total Variation Distance&   $D_{TVD}\big(p(|\theta)||g()\big)$  & $\int |p(X|\theta) - g(X) | dX $\\
    alpha&   $D_{\alpha}\big(p(|\theta)||g()\big)$&$\frac{1}{\alpha(1-\alpha)}\Big(1-\int p(X|\theta)^\alpha g(X)^{1-\alpha} dX \Big)$ 
    \end{tabular}
    \caption{The divergences used as losses in this work.}
\label{tab:divergences}
\end{table}

\begin{figure}[h]

\begin{subfigure}{0.5\textwidth}
\includegraphics[width=0.9\linewidth, height=5cm]{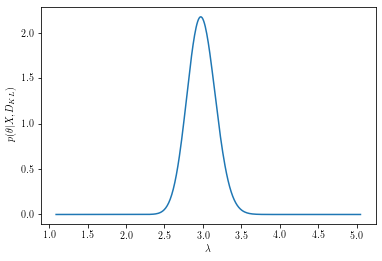} 
\caption{Bayes update $p(\theta|X,d_{KL})$ for a well-specified Poisson model.}
\label{fig:subim1}
\end{subfigure}
\begin{subfigure}{0.5\textwidth}
\includegraphics[width=0.9\linewidth, height=5cm]{figures/poisson_KL_wellspec.png}
\caption{Bayes update $p(\theta|X,d_{KL})$ for a misspecified Poisson model.}
\label{fig:subim2}
\end{subfigure}

\begin{subfigure}{0.5\textwidth}
\includegraphics[width=0.9\linewidth, height=5cm]{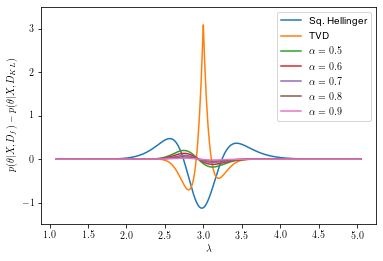} 
\caption{Difference between Bayes update $p(\theta|X,d_{KL})$ and $p(\theta|X,d_{f})$ using divergence $d_f$ for a well-specified Poisson model.}
\label{fig:subim1}
\end{subfigure}
\begin{subfigure}{0.5\textwidth}
\includegraphics[width=0.9\linewidth, height=5cm]{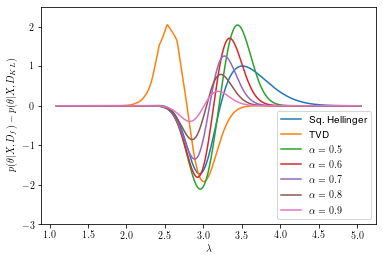}
\caption{Difference between Bayes update $p(\theta|X,d_{KL})$ and $p(\theta|X,d_{f})$ using divergence $d_f$ for a misspecified Poisson model.}
\label{fig:subim2}
\end{subfigure}
\caption{A comparison of belief updates using different divergences as losses for a Poisson statistical model, with observed data generated from $\mathrm{Pois}(3)$ in the well-specified case and $\mathrm{NB}(10,0.8)$ in the misspecified case. }
\label{fig:poisson_updates}

\end{figure}

The classifier used was a logistic regression model, trained on summary statistics judged to be appropriate for each problem. Inference for the classifier was performed using three difference methods, which were compared. The first was a lasso-regularised method using cross-validation to tune the regularisation strength; the second was a Bayesian method using Empirical Bayes to perform a Laplace approximation for the classification parameters, and the third was a Bayesian method using Variational Bayes for the classifier inference.

The classifier methods were compared with methods using the statistical model likelihood $p(|\theta)$ and a generative nonparametric model to model the data generative process $g()$. In the case of unsupervised modelling, a KDE was used to describe $g()$, and for the supervised modelling a Gaussian Process regression was used.  Acquisitions for the generative modelling method were performed both on a dense grid and using BayesOpt, to assess the influence of BayesOpt on the quality of the approximation.

\section{Examples}\label{sec:examples}

In this section, we present three examples based on Poisson, Gaussian and linear regression statistical models. Situations are considered in which the statistical model is well-specified and poorly specified. The observed data are simulated from a known distribution $q_{\mathcal{T}}()$ representing the true generative process. Consequently, it is possible to define the belief distribution $p(\theta|X_\text{obs},\mathcal{T},f)$ using the approximations in Equation \eqref{eqn:forward_obs} as an ideal belief update to compare against:

\begin{align}
    p(\theta|X_\text{obs},\mathcal{T},f) \propto \exp\Big(-n_\text{obs} D_f\big(p(|\theta)||q_{\mathcal{T}}()\big)\Big) p(\theta)
\end{align}

Example belief updates for a Poisson model are shown in Figure \ref{fig:poisson_updates}: we see the difference between the standard Bayes updates and general divergence updates conditional on the true observational process $q_{\mathcal{T}}()$. With the exception of the TVD, the non-KL updates deviate further from a classical Bayesian update in the misspecified case, with the heavier tails of the negative binomial distribution clearly influencing the inference. As expected, the deviations for alpha divergences updates from standard Bayes updates also grow smaller with increasing $\alpha$.

We then use the Jensen-Shannon divergence (JSD) to evaluate the distance between the belief distribution $p(\theta|X_\text{obs},g,f)$ from each of the inference methods against $p(\theta|X_\text{obs},\mathcal{T},f)$. Experiments were repeated over fifty random seeds, and the results presented are the mean of the JSDs over the repetitions. Uniform priors $p(\theta)$ were used for all examples in this article.

For every example in this article, updates were performed conditioned on 90 observed data points drawn from the true distribution $q_{\mathcal{T}}(X)$. 81 simulated data points were used per loss evaluation to enable 10-fold cross-validation on the observed data.

The JSD used here is defined as the square root of the mean of the KL divergence and the reverse KL divergence between the two distributions. It is bounded above by a value of $\sqrt{\ln(2)}\approx 0.8325$. For the KL divergence, the generative distribution $g(X)$ becomes absorbed within the normalisation and does not contribute to the functional form of the belief distribution, and as such the methods using the true likelihood are expected to return results equivalent to $p(\theta|X_\text{obs},\mathcal{T},f)$, conditional on the success of the BayesOpt procedure. The BayesOpt procedures used an upper-confidence bound acquisition function with a relatively large exploration parameter $\beta=5$ for all inference methods to ensure reasonable global exploration of the parameter space, 100 total acquisition steps, and an additive Matern32 kernel and constant kernel. 

\subsection{Poisson}

In this case we consider inference for a Poisson statistical model $X\sim \mathrm{Pois}(\lambda)$, with data $X$ and parameter space $\lambda$ each of one dimension. We consider a well-specified case and misspecified case, in which the observed data are drawn from a Poisson distribution and a negative binomial distribution, respectively.

The features used in the classifier were simply the untransformed data samples $X$ and a constant. The generative modelling for the likelihood-based method was performed using a KDE with a Gaussian kernel. The generative grid approach and evaluation against the true update used a grid of 1001 points.

\subsubsection{Well-specified Poisson}

In this section, we consider a Poisson statistical model in the well-specified situation, with the observed data $X_\text{obs}$ being drawn from a Poisson model with rate parameter of 3, i.e.~$X_\text{obs}\sim \mathrm{Pois}(3)$. The results from the experiments are presented in Table \ref{tab:poisson_results}. We see that, for the TVD, the squared Hellinger distances, and alpha divergences with $\alpha \leq 0.8$, the classifier-based inference outperforms the generative inference with access to the model likelihood. The likelihood-based inference performs best with the KL divergence and the alpha divergence with $\alpha=0.9$, for which the influence of the KDE generative modelling is minimal. The Variational Bayes classifier appear to perform consistently better than the other classifiers in this context.

\begin{table}
\begin{tabular}{ c|c|c c c c c }
True $q_{\mathcal{T}}()$ & Divergence & CV classifier & EB classifier & VB classifier & Gen BayesOpt & Gen grid \\
\hline
\multirow{8}{5em}{$\mathrm{Pois}(3)$} & TVD & 0.1965 & $\mathbf{0.1927}$ & 0.1981 & 0.2723 &  0.2788 \\
& Sq. Hellinger & 0.1314 & 0.1282 & $\mathbf{0.1199}$ & 0.1759 & 0.1781 \\
& $\alpha=0.5$ & 0.2250 & 0.2184 & $\mathbf{0.1916}$ & 0.3215 & 0.3260 \\
& $\alpha=0.6$ & 0.1738 & 0.1690 & $\mathbf{0.1342}$ & 0.2263 & 0.2329 \\
& $\alpha=0.7$ & 0.1396 & 0.1335 & $\mathbf{0.09553}$ & 0.1510 & 0.1589 \\
& $\alpha=0.8$ & 0.1198 & 0.1095 & $\mathbf{0.06939}$ & 0.09039 &  0.09887 \\
& $\alpha=0.9$ & 0.1068 & 0.09615 & 0.05690 & $\mathbf{0.04085}$ & 0.04729 \\
& KL & 0.09854 & 0.09191 & 0.05922 & 0.01530 & $\mathbf{0.002426}$ \\
\hline
\multirow{8}{5em}{$\mathrm{NB}(10,0.8)$} & TVD & 0.3212 & $\mathbf{0.3154}$ & 0.3161 & 0.3496 & 0.3353 \\
& Sq. Hellinger & $\mathbf{0.1029}$ & 0.1235 & 0.1141 & 0.1163 & 0.1182 \\
& $\alpha=0.5$ & $\mathbf{0.1957}$ & 0.2250 & 0.2168 & 0.2133 & 0.2162 \\
& $\alpha=0.6$ & $\mathbf{0.1340}$ & 0.1684 & 0.1580 & 0.14863 & 0.1524 \\
& $\alpha=0.7$ & $\mathbf{0.09217}$ & 0.1302 & 0.1200 & 0.09806 & 0.1019 \\
& $\alpha=0.8$ & 0.07452 & 0.10482 & 0.09742 & $\mathbf{0.05792}$ & 0.06126 \\
& $\alpha=0.9$ & 0.07674 & 0.08940 & 0.08694 & $\mathbf{0.02576}$ & 0.02791 \\
& KL & 0.08716 & 0.08135 & 0.08630 & 0.01229 & $\mathbf{0.00005307}$ \\
\end{tabular}\caption{Results for a Poisson-distributed statistical model $\mathrm{Pois}(\lambda)$, with observed data drawn from the true data generating process $q_{\mathcal{T}}()$ indicated. Data are mean JSDs averaged over fifty seeds, between the generalised belief distributions generated by each inference method and the belief update using the true unknown generative process, using each divergence as a loss function.}
\label{tab:poisson_results}
\end{table}

\subsubsection{Misspecified Poisson}

In this section, we consider a Poisson statistical model in the misspecified situation, with the observed data $X_\text{obs}$ begin drawn from a negative binomial model with stopping parameter 10 and success probability of 0.8, i.e.~$X_\text{obs}\sim \mathrm{NB}(10,0.8)$. The results from the experiments are presented in Table \ref{tab:poisson_results}.

We see that the classifier-based inference performs best for the TVD, Squared Hellinger divergence and alpha divergences with $\alpha \leq 0.7$, while assuming the statistical likelihood and performing generative modelling of $g()$ performed best for the KL and alpha divergences with $\alpha \geq 0.8$

\subsection{Gaussian}

In this section, we consider a Gaussian statistical model with unknown mean $\mu$ and variance $\sigma^2$, giving a one dimensional data space and two-dimensional parameter space, i.e.~$X\sim \mathcal{N}(\mu,\sigma^2)$. Inference was performed over the mean and log transform of the variance to increase the stability of the BayesOpt. We consider well-specified and misspecified contexts, in which the observed data are drawn from a Gaussian distribution and a Laplace distribution, respectively.

The data transformations used in the classifier were $X$ and $|X-\bar{X}|$, where $\bar{X}$ is the mean of the sample. The generative modelling for the likelihood-based method was performed using a KDE with a Gaussian kernel. The generative grid approach and evaluation against the true update used a two-dimensional grid with 101 points in each dimension.

\subsubsection{Well-specified}

We consider the well-specified Gaussian case here, in which the observed data $X_\text{obs}$ are drawn from a Normal distribution with unit mean and variance, i.e. $X_\text{obs}\sim \mathcal{N}(1,1)$.

The results of the experiments are presented in Table \ref{tab:gaussian_results}. We see from the results that the classifier-based inference performs best for TVD, Squared Hellinger divergence, and for alpha divergences with $\alpha \leq 0.6$. The generative modelling on a grid is very slightly more successful for the alpha divergence with $\alpha=0.7$, and clearly more successful with $\alpha \geq 0.8$ and the KL.

We see a noticeable difference between the methods relying on BayesOpt and the grid-based generative approach, suggesting that the use of fewer acquisitions by BayesOpt may be considered in a trade-off with accuracy. The classifier-based approaches outperform the generative BayesOpt approach for every divergence considered, including the KL.

\begin{table}
\begin{tabular}{ c|c|c c c c c }
True $q_{\mathcal{T}}()$ & Divergence & CV classifier & EB classifier & VB classifier & Gen BayesOpt & Gen grid \\
\hline
\multirow{8}{6em}{$\mathcal{N}(1,1)$} & TVD & 0.4895 & 0.4964 & $\mathbf{0.4686}$ & 0.6356 & 0.6780 \\
& Sq. Hellinger & 0.3298 & $\mathbf{0.3242}$ & 0.3334 & 0.4980 & 0.4652 \\
& $\alpha=0.5$ & 0.5376 & $\mathbf{0.5344}$ & 0.5448 & 0.7072 & 0.7064 \\
& $\alpha=0.6$ & $\mathbf{0.4778}$ & 0.4807 & 0.4828 & 0.6696 & 0.5831 \\
& $\alpha=0.7$ & 0.4358 & 0.4441 & 0.4419 & 0.6550 &  $\mathbf{0.4336}$ \\
& $\alpha=0.8$ & 0.4099 & 0.4230 & 0.4205 & 0.6669 & $\mathbf{0.2774}$ \\
& $\alpha=0.9$ & 0.3973 & 0.4145 &   0.4147 & 0.6943 & $\mathbf{0.1299}$ \\
& KL & 0.3956 & 0.4162 & 0.4209 & 0.7253 & $\mathbf{0.0005892}$ \\
\hline
\multirow{8}{6em}{$\mathrm{Laplace}(1,1)$} & TVD & $\mathbf{0.2991}$ & 0.3106&  0.3111 & 0.5066 & 0.5833 \\
& Sq. Hellinger & 0.3759 & $\mathbf{0.3680}$ & 0.3747 & 0.5221 & 0.5793 \\
& $\alpha=0.5$ & 0.59928 & $\mathbf{0.5923}$ & 0.5964 & 0.7248 & 0.7824 \\
& $\alpha=0.6$ & 0.5201 & $\mathbf{0.5161}$ & 0.5205 & 0.6853 &  0.6749 \\
& $\alpha=0.7$ & 0.4647 & $\mathbf{0.4623}$ & 0.4653 & 0.6447 & 0.5064 \\
& $\alpha=0.8$ & 0.4313 &  0.4325 & 0.4305 & 0.6206 & $\mathbf{0.3238}$ \\
& $\alpha=0.9$ & 0.4148 & 0.4225 &  0.4160 & 0.6225 & $\mathbf{0.1655}$ \\
& KL & 0.4135 & 0.4275 & 0.4182 & 0.6507 & $\mathbf{0.04343}$ \\
\end{tabular}\caption{Results for a Gaussian-distributed statistical model $\mathcal{N}(\mu,\sigma^2)$, with observed data drawn from the true data generating process $q_{\mathcal{T}}()$ indicated. Data are mean JSDs averaged over fifty seeds, between the generalised belief distributions generated by each inference method and the belief update using the true unknown generative process, using each divergence as a loss function.}
\label{tab:gaussian_results}
\end{table}

\subsubsection{Misspecified}

We consider the misspecified Gaussian case here, in which the observed data $X_{obs}$ are drawn from a Laplace distribution with unit location and scale parameter, i.e. $X_{obs}\sim \mathrm{Laplace}(1,1)$. 

The results of the experiments are presented in Table \ref{tab:gaussian_results}. We see the classifier-based inference outperforms all the generative likelihood-based methods for the TVD, Squared Helliner, and alpha divergence with $\alpha \leq 0.7$, while the generative grid approach performs best for the KL and $\alpha \geq 0.8$. We again see a significant difference between the generative BayesOpt approach and grid approach, especially for larger values of $\alpha$: all the classifier-based approaches outperform the generative BayesOpt approach for every divergence considered.

\subsection{Linear Regression}

In this section we consider inference in a supervised context, with a linear regression statistical model. The model assumes Gaussian noise, i.e. $y\sim\mathcal{N}(\beta_0+\beta_1 X,\sigma^2)$. The model has one response variable $y$ and one covariate $X$, and three parameters $beta_0$, $beta_1$ and $\sigma^2$ describing the regression intercept, regression slope and noise variance, respectively. Inference was performed directly over $\beta_0$, $\beta_1$ and the log transform of $\sigma^2$ to increase the stability of the BayesOpt procedure.

In this section, we consider both a well-specified and misspecified case, where the observed data are drawn from a linear model with Gaussian noise and Student t-distributed noise, respectively.

The data features provided to the classifier were $y$, $(y-\bar{y})^2$, $(y-\bar{y})^{4}/{\sigma_y^4}$, $X$, $(X-\bar{X})^2$ and $(y-\bar{Y})(X-\bar{X})$, where $\bar{X}$ and $\bar{y}$ were the mean of the sample covariates and responses, respectively, and $\sigma_y$ is the sample standard deviation of the responses. The generative model $g(y|X)$ used for the likelihood-based method was Gaussian Process regression with a Matern32 kernel and constant kernel. The generative grid approach and evaluation against the true belief update used a three-dimensional grid with 51 points in each dimension.

\subsubsection{Well-specified}

In this section, we present experiments for a well-specified linear regression, with observed data $y_{obs}, X_{obs}$being drawn from a linear model with Gaussian-distributed noise and $\beta_0 = 0$, $\beta_1=-1$ and $\sigma=0.5$, i.e. $y_{obs}\sim\mathcal{N}(-X_{obs},0.5^2)$. Covariates $X_{obs}$ were drawn from a standard normal distribution, i.e. $X_{obs}\sim\mathcal{N}(0,1)$.

Results are presented in Table \ref{tab:regression_results}. We observe that the classifier-based inference performs best for the TVD, Squared Hellinger and alpha divergences with $\alpha \leq 0.7$, while the generative grid approach performs best for $\alpha \geq 0.8$ and the KL. The cross-validated classifier approach performs consistently better than the other classifier inference methods. We see a very significant difference bettwen the generative BayesOpt method and generative grid approach, especially for large values of $\alpha$, suggesting that the BayesOpt procedure is having a significant effect on results.

\begin{table}
\begin{tabular}{ c|c|c c c c c }
True $q_{\mathcal{T}}()$ & Divergence & CV classifier & EB classifier & VB classifier & Gen BayesOpt & Gen grid \\
\hline
\multirowcell{8}{$\mathcal{N}(\mu=-X_{obs},$\\$ \quad  \sigma^2=0.5^2)$} & TVD & 0.5951 & 0.5672 & $\mathbf{0.5073}$ & 0.7790 & 0.7543 \\
& Sq. Hellinger & $\mathbf{0.5039}$ & 0.5634 & 0.5241 & 0.6058 & 0.6222 \\
& $\alpha=0.5$ & $\mathbf{0.6632}$ & 0.7312 & 0.6976 & 0.7606 & 0.7651 \\
& $\alpha=0.6$ & $\mathbf{0.5921}$ & 0.6911 & 0.6361 & 0.7483 & 0.7162 \\
& $\alpha=0.7$ & $\mathbf{0.5366}$ & 0.6612 & 0.5904 & 0.7583 & 0.6021 \\
& $\alpha=0.8$ & 0.4941 & 0.6387 & 0.5595 & 0.7774 & $\mathbf{0.4356}$ \\
& $\alpha=0.9$ & 0.4684 & 0.6262 & 0.5474 & 0.7966 & $\mathbf{0.2370}$ \\
& KL & 0.4596 & 0.6210 & 0.5464 & 0.8087 & $\mathbf{0.04124}$ \\
\hline
\multirowcell{8}{$\mathbf{t}(\mu=-X_{obs},$\\$  \sigma=0.5,$\\$  \nu=3)$} & TVD & $\mathbf{0.5046}$ & 0.5301 & 0.5011 & 0.7280 & 0.7254 \\
& Sq. Hellinger & $\mathbf{0.5256}$ & 0.5777 & 0.5670 & 0.5796 & 0.6810 \\
& $\alpha=0.5$ & $\mathbf{0.6964}$ & 0.7455 & 0.7420 & 0.7661 & 0.7928 \\
& $\alpha=0.6$ & $\mathbf{0.6155}$ & 0.6952 & 0.6859 & 0.7525 & 0.7672 \\
& $\alpha=0.7$ & $\mathbf{0.5488}$ & 0.6589 & 0.6375 & 0.7652 & 0.6838 \\
& $\alpha=0.8$ & $\mathbf{0.5112}$ & 0.6318 &  0.6080 & 0.7788 & 0.5308 \\
& $\alpha=0.9$ & 0.5100 & 0.6159 & 0.6046 & 0.7916 & $\mathbf{0.3490}$ \\
& KL & 0.5277 & 0.6098 &  0.6155 & 0.8010 & $\mathbf{0.1549}$ \\
\end{tabular}\caption{Results for a linear regression statistical model with Gaussian noise $\mathcal{N}(\beta_0+\beta_1 X_{obs},\sigma^2)$, with observed data$y_{obs}$  drawn from the true data generating process $q_{\mathcal{T}}()$ indicated. Data are mean JSDs averaged over fifty seeds, between the generalised belief distributions generated by each inference method and the belief update using the true unknown generative process, using each divergence as a loss function.}
\label{tab:regression_results}
\end{table}

\subsubsection{Misspecified}

In this section, we present experiments for a misspecified linear regression, with observed data $y_{obs}, X_{obs}$being drawn from a linear model with Student t-distributed noise with 3 degrees of freedom, and $\beta_0 = 0$, $\beta_1=-1$ and $\sigma=0.5$, i.e. $y_{obs}\sim\mathbf{t}(\mu=-X_{obs},\sigma=0.5,\nu=3)$. Covariates $X_{obs}$ were drawn from a standard normal distribution, i.e. $X_{obs}\sim\mathcal{N}(0,1)$.

Results are presented in Table \ref{tab:regression_results}. The classifier-based inference performs best for TVD, Square Hellinger and alpha divergences with $\alpha \leq 0.8$, while the generative grid approach performs best for the KL and the alpha divergence with $\alpha=0.9$. As in the well-specified case, the cross-validated classifier consistently performs better than the other classifiers, and the generative approach with BayesOpt performs the worst of all the methods, especially for the KL and larger values of $\alpha$. We observe that the JSD for the KL divergence update using generative modelling and a grid deviates significantly from zero, possibly from the ratio truncation procedure or numerical instability giving the generative modelling influence over the belief updates.

\section{Conclusion}\label{sec:conclusion}

From the results presented in this work, we conclude that classifiers represent a useful method for constructing generalised belief updates based on $f$-divergences for interpretable statistical models, often matching or even substantially outperforming methods that use the correct model likelihood and generative modelling for $g()$. The use of discriminative classifiers represents a practical inference method of potential use in situations where pathological over-concentration of the posterior has previously been observed due to mismatch between data and the model \cite{yang2007fair,yang2018bayesian}.

The results in this article divide the divergences fairly clearly into two sets: the classifier-based inference performs best for Squared Hellinger, TVD, and alpha divergences with $\alpha$ closer to 0.5 than 1, whereas the likelihood-based inference with generative modelling for $g()$ performs best for the KL divergence and alpha divergences with $\alpha$ close to 1. This makes sense considering the definition of the divergences, as the KL and alpha divergences with larger $\alpha$ put less emphasis on the generative model $g()$ and more on the model likelihood $f(|\theta)$, so methods that assume the model likelihood do well, even if the generative modelling for $g()$ is challenging.

The two sets of divergences themselves correspond to different judgements to be made by the statistician. The TVD, Squared Hellinger, and alpha divergences with smaller values of $\alpha$ all give less influence to the tails of the distribution of the data, producing a belief update that is robust to outliers. They are an appropriate choice when the statisticians is primarily interested in bulk properties of data away from the tails, or the tails are suspected to be misspecified. By contrast, the KL divergence and alpha divergences with larger values of $\alpha$ give larger influence to the tails of the data distribution, so are appropriate when the tail behaviour is important and assumed to be well-specified.

Use of the KL divergence corresponds to an exact Bayes update and can be performed without any modelling of $g()$. Given the results presented here, we suggest that discriminative modelling is preferable for generalised Bayes updates using all of the divergences considered that promote strong robustness to tail behaviour. Generative modelling of $g()$ appears only appropriate for updates based on alpha divergences with large values of $\alpha$, i.e. something close to a traditional Bayesian update with weak damping on the tails of the data.

It is possible that the success of the BayesOpt procedure is having an effect on the results presented here, given that we frequently see a significant difference between the generative method using a grid and using BayesOpt. The classifier-based methods consistently outperform the generative methods also using BayesOpt, which suggests that the discriminative ratio estimation method is more accurate with the same number of limited acquisitions. It is not clear from the results whether the differences in performance between BayesOpt-based discriminative methods and the grid-based generative methods are due to the influence of BayesOpt or the different density estimation methods. Given that the ratio estimation using classifiers is more computationally expensive than the likelihood evaluations for generative modelling, then this comparison is of pragmatic importance.

It would be reasonable for further work to explore different acquisition methods, given the apparent influence of standard BayesOpt on the inference results. It is possible that developing acquisitions specific to each divergence may help, or that alternative acquisition methods will stabilise the belief distribution approximation.

None of the three classifiers used in this work were universally more successful than the others, although some were consistently more accurate for specific combinations of statistical model and true data generating process. It would be instructive to explore the effects of different methods for the classifier, which may also have implications for automatic selection of summary statistics for a given problem through regularisation. It would also be productive to possibly avoid the use of summary statistics entirely through the use of neural network methods or nonparametric Gaussian Process Classification.

Finally, it would be of interest to consider divergences not evaluated in this article. Other f-divergences are of possible interest as the loss defining a generalised belief update, including the Jensen-Shannon divergence or the reverse defined KL or alpha divergences. The choice of divergence is a modelling decision depending on the subjective judgement of the statistician, so a more expansive analysis of the appropriate computational approximations for each would be of interest.

\bibliographystyle{unsrt}  
\bibliography{references}  

\begin{thebibliography}{10}

\bibitem{bernardo2009bayesian}
Jos{\'e}~M Bernardo and Adrian~FM Smith.
\newblock {\em Bayesian theory}, volume 405.
\newblock John Wiley \& Sons, 2009.

\bibitem{walker2013bayesian}
Stephen~G Walker.
\newblock Bayesian inference with misspecified models.
\newblock {\em Journal of Statistical Planning and Inference},
  143(10):1621--1633, 2013.

\bibitem{yang2007fair}
Ziheng Yang.
\newblock Fair-balance paradox, star-tree paradox, and bayesian phylogenetics.
\newblock {\em Molecular biology and evolution}, 24(8):1639--1655, 2007.

\bibitem{yang2018bayesian}
Ziheng Yang and Tianqi Zhu.
\newblock Bayesian selection of misspecified models is overconfident and may
  cause spurious posterior probabilities for phylogenetic trees.
\newblock {\em Proceedings of the National Academy of Sciences},
  115(8):1854--1859, 2018.

\bibitem{bissiri2016general}
Pier~Giovanni Bissiri, Chris~C Holmes, and Stephen~G Walker.
\newblock A general framework for updating belief distributions.
\newblock {\em Journal of the Royal Statistical Society: Series B (Statistical
  Methodology)}, 78(5):1103--1130, 2016.

\bibitem{thomas2020split}
Owen Thomas, Henri Pesonen, Raquel S{\'a}-Le{\~a}o, Herm{\'\i}nia de~Lencastre,
  Samuel Kaski, and Jukka Corander.
\newblock Split-bolfi for for misspecification-robust likelihood free inference
  in high dimensions.
\newblock {\em arXiv preprint arXiv:2002.09377}, 2020.

\bibitem{jewson2018principles}
Jack Jewson, Jim~Q Smith, and Chris Holmes.
\newblock Principles of bayesian inference using general divergence criteria.
\newblock {\em Entropy}, 20(6):442, 2018.

\bibitem{sugiyama2012density}
Masashi Sugiyama, Taiji Suzuki, and Takafumi Kanamori.
\newblock {\em Density ratio estimation in machine learning}.
\newblock Cambridge University Press, 2012.

\bibitem{sisson2018handbook}
Scott~A Sisson, Yanan Fan, and Mark Beaumont.
\newblock {\em Handbook of approximate Bayesian computation}.
\newblock Chapman and Hall/CRC, 2018.

\bibitem{gutmann2018likelihood}
Michael~U Gutmann, Ritabrata Dutta, Samuel Kaski, and Jukka Corander.
\newblock Likelihood-free inference via classification.
\newblock {\em Statistics and Computing}, 28(2):411--425, 2018.

\bibitem{thomas2016likelihood}
Owen Thomas, Ritabrata Dutta, Jukka Corander, Samuel Kaski, and Michael~U
  Gutmann.
\newblock Likelihood-free inference by ratio estimation.
\newblock {\em arXiv preprint arXiv:1611.10242}, 2016.

\bibitem{hermans2019likelihood}
Joeri Hermans, Volodimir Begy, and Gilles Louppe.
\newblock Likelihood-free mcmc with amortized approximate likelihood ratios.
\newblock {\em stat}, 1050:1, 2019.

\bibitem{cranmer2015approximating}
Kyle Cranmer, Juan Pavez, and Gilles Louppe.
\newblock Approximating likelihood ratios with calibrated discriminative
  classifiers.
\newblock {\em arXiv preprint arXiv:1506.02169}, 2015.

\bibitem{tran2017hierarchical}
Dustin Tran, Rajesh Ranganath, and David Blei.
\newblock Hierarchical implicit models and likelihood-free variational
  inference.
\newblock In {\em Advances in Neural Information Processing Systems}, pages
  5523--5533, 2017.

\bibitem{louppe2019adversarial}
Gilles Louppe, Joeri Hermans, and Kyle Cranmer.
\newblock Adversarial variational optimization of non-differentiable
  simulators.
\newblock In {\em The 22nd International Conference on Artificial Intelligence
  and Statistics}, pages 1438--1447, 2019.

\bibitem{thomas2019diagnosing}
Owen Thomas and Jukka Corander.
\newblock Diagnosing model misspecification and performing generalized bayes'
  updates via probabilistic classifiers.
\newblock {\em arXiv preprint arXiv:1912.05810}, 2019.

\bibitem{goodfellow2014generative}
Ian Goodfellow, Jean Pouget-Abadie, Mehdi Mirza, Bing Xu, David Warde-Farley,
  Sherjil Ozair, Aaron Courville, and Yoshua Bengio.
\newblock Generative adversarial nets.
\newblock In {\em Advances in neural information processing systems}, pages
  2672--2680, 2014.

\bibitem{nowozin2016f}
Sebastian Nowozin, Botond Cseke, and Ryota Tomioka.
\newblock f-gan: Training generative neural samplers using variational
  divergence minimization.
\newblock In {\em Advances in neural information processing systems}, pages
  271--279, 2016.

\bibitem{williams2006gaussian}
Christopher~KI Williams and Carl~Edward Rasmussen.
\newblock {\em Gaussian processes for machine learning}, volume~2.
\newblock MIT press Cambridge, MA, 2006.

\bibitem{shahriari2015taking}
Bobak Shahriari, Kevin Swersky, Ziyu Wang, Ryan~P Adams, and Nando De~Freitas.
\newblock Taking the human out of the loop: A review of bayesian optimization.
\newblock {\em Proceedings of the IEEE}, 104(1):148--175, 2015.

\bibitem{gutmann2016bayesian}
Michael~U Gutmann and Jukka Corander.
\newblock Bayesian optimization for likelihood-free inference of
  simulator-based statistical models.
\newblock {\em The Journal of Machine Learning Research}, 17(1):4256--4302,
  2016.

\end{thebibliography}



\end{document}